\documentclass[12pt]{article}
\usepackage[dvips]{epsfig}
\def\be{\begin{equation}}
\def\bea{\begin{eqnarray}}
\def\ee{\end{equation}}
\def\eea{\end{eqnarray}}

\def\d{\partial}

\begin{document}
\begin{flushright}
OHSTPY-HEP-T-99-015\\
hep-th/9910012
\end{flushright}
\vspace{20mm}
\begin{center}
{\LARGE  Supersymmetry and DLCQ \footnote{ TJNAF Workshop in honor of
Stan Brodsy's 60th Birthday, University of Georgia, September 17, 1999}}
\\
\vspace{20mm}
{\bf  The SDLCQ Collaboration \footnote {Francesco
Antonuccio, I. Filippov, P. Haney, O. Lunin,
S.Pinsky, U. Trittmann, Department of Physics, The Ohio State University,
Columbus, OH 43210, and J. Hiller University of Minnesota-Duluth MI 55812}}\\
\vspace{4mm}
\end{center}
\vspace{10mm}

\begin{abstract}
In this talk we describe the application of discrete light cone quantization
(DLCQ) to supersymmetric field theories. We find that it is possible to
formulate DLCQ so that supersymmetry is exactly preserved in the discrete
approximation and call this formulation of DLCQ, SDLCQ.  It combines the
power of DLCQ with all of the beauty of supersymmetry. We have applied
SDLCQ to several interesting supersymmetric theories and
discussed  zero modes, vacuum degeneracy, massless states, mass gaps, and
theories in higher dimensions. Most recently we have used it to discuss the
Maldacena conjecture.
\end{abstract}
\newpage

\baselineskip .25in

\section{Introduction}

In the last decade there have been significant improvements in our
understanding of gauge theories and important breakthroughs in the
nonperturbative description of supersymmetric gauge theories
\cite{seiberg,seibergwitten}. In the last few year various relations between
string theory, brane theory and gauge fields \cite{givkut,agmoo} have also
emerged. While these developments give us some insight into strongly coupled
gauge theories
\cite{seibergwitten}, they do not offer a direct method for non-perturbative
calculations. In this talk we discuss some recent developments in the
light cone quantization approach to non-perturbative problems. We have found
that these methods have the potential to expand our understanding of strongly
coupled gauge theories in directions not previously available.

The original idea of light cone quantization was formulated half of a century
ago
\cite{Dirac}, but apart from several technical clarifications \cite{MaYam}
it remained mostly undeveloped. The first change came in the mid 80's when
the Discrete Light Cone
Quantization (DLCQ) was suggested as a practical way for calculating the
masses and wavefunctions of hadrons \cite{BP85}. Although the direct
application of the method to realistic problems meets some difficulties (for
review see \cite{BPP}), DLCQ has been successful in studying various two
dimensional models. Given the importance of supersymmetric theories, it is not
surprising that light cone quantization was ultimately applied to such models
\cite{kut93,bdk93,kub94}.  In these early works the mass spectrum was
shown to be supersymmetric in the continuum and a great deal of information
about the properties of bound states in supersymmetric theories was extracted.
However the straightforward application of DLCQ to the supersymmetric systems
had one disadvantage: the supersymmetry was lost in the discrete formulation.
The way to solve this problem was suggested in \cite{sakai}, where the
alternative formulation of DLCQ was introduced. Namely it was noted that since
the supercharge is the "square root" of the Hamiltonian one can define a new
DLCQ procedure based on the supercharge. We will discuss this formulation
(called SDLCQ) in this talk.

\section{ \bf Supersymmetric Yang--Mills Theory in the Light--Cone Gauge.}
\label{ChLCGauge}

We will consider here the bound state problem for various supersymmetric
matrix models in two dimensions. These models may be
constructed by dimensional reduction of supersymmetric Yang--Mills theory in
higher dimensions.
Before we begin the a discussion of bound state problem
for a specific systems it is worthwhile to summarize some basic
ideas of Discrete Light Cone Quantization, for a complete review see
\cite{BPP}.

Let us consider general relativistic systems in two dimensions. In the usual
canonical quantization of such systems one imposes certain commutation
relations between coordinates and momenta at equal time. However, as was
pointed out by Dirac long ago \cite{Dirac}, this is not the only possibility.
Another scheme of quantization treats the light like coordinate
$x^+=\frac{1}{\sqrt{2}}(x^0+x^1)$ as a
new time and then the system is quantized canonically. This scheme (called
light cone quantization) has both positive and negative aspects. The main
disadvantage of light cone quantization is the presence of constraints.
Even  systems as simple as free bosonic field has one: from the action
\be
S=\int d^2 x \d_+\phi \d_-\phi
\ee
one can derive the constraint relating coordinate and momentum:
\be
\pi=\d_-\phi.
\ee
For more complicated systems the constraints are also present and in general
they are hard to resolve.

The main advantage of the light cone is the decoupling of positive and
negative momentum modes. This property is crucial for DLCQ. In the
Discrete Light Cone Quantization one considers the theory on the finite circle
along the $x^-$ axis: $-L<x^-<L$. Then all the momenta become quantized and
the integer number measuring the total momentum in terms of "elementary
momentum" is called the harmonic resolution $K$. Due to the decoupling
property one
may work only in the sector with positive momenta and as a result there is
a finite
number of states for any finite value of resolution. Of course the full
quantum field theory in the continuum corresponds to the limit
$L\rightarrow\infty$, and in this limit the elementary bit of momentum goes to
zero, as the harmonic resolution goes to infinity and the infinite number
of degrees of freedom are restored. It is believed that the "quantum
mechanical" approximation is suitable for describing the lowest states in the
spectrum.
Note that the problem of constraints in DLCQ is a quantum mechanical one and
thus it is easier to solve. Usually this problem can be reformulated in terms
of zero modes and the solution can be found for any value of the resolution.

DLCQ is mainly used to solve the bound state problem and we will
formulate this problem for a general two dimensional theory. The theory in the
continuum has full Poincare symmetry, thus the states are naturally
labeled by the eigenvalues of Casimir operators of the Poincare algebra. One
such Casimir is the mass operator: $M^2=P^\mu P_\mu$. Another Casimir is
related to the spin of the particle and is generally not used. After
compactifying
the $x^-$ direction one looses Lorentz symmetry, but not the translational
invariance in $x^+$ and $x^-$ directions. Thus $P^+$ and $P^-$ are still
conserved charges, and the states are characterized by both $P^+$ and $P^-$.
If we consider DLCQ as an approximation to the continuum theory, we anticipate
that in the limit of infinite harmonic resolution (or $L\rightarrow\infty$)
the full Poincare symmetry is restored. Thus the aim would be to study the
value of $M^2$
as function of $K$ and to extrapolate the results to the $K=\infty$.

The usual way to define $M^2$ in DLCQ is based on separate calculation of
$P^+$ and $P^-$ in matrix form and then bringing them together:
\be
M^2=2P^+P^-.
\ee
Usually one works in the sector with fixed $P^+$, but the calculation of
light cone Hamiltonian $P^-$ is a nontrivial problem. Important
simplifications occur
for supersymmetric theories \cite{sakai}.

Supersymmetry is the only
nontrivial extension of Poincare algebra compatible with the existence of an S
matrix \cite{WessBag}. Namely in addition to usual bosonic generators of
symmetries, fermionic ones are allowed and the full (super)algebra in
two dimensions reads:
\bea\label{SUSYalg}
\{Q^I_\alpha,Q^J_\beta\}&=&2\delta^{IJ}\gamma^\mu_{\alpha\beta}P_\mu+
  \varepsilon_{\alpha\beta}Z^{IJ},\\
\left[ P_\mu,P_\nu \right]&=&0,\qquad
\left[P_\mu,Q^I_\alpha \right]=0.
\eea
In this expression $\varepsilon$ is the antisymmetric $2\times 2$ matrix,
$\varepsilon_{12}=1$ and $Z^{IJ}$ is the set of c--numbers called central
charges. In this talk we will put them equal to zero.
It is convenient to choose two dimensional gamma matrices in the form:
$\gamma^0=\sigma^2$, $\gamma^1=i\sigma^1$, then one can rewrite
(\ref{SUSYalg}) in terms of light cone components:
\bea
\{Q^+_I,Q^+_J\}&=&2\sqrt{2}\delta^{IJ}P^+,\\
\{Q^-_I,Q^-_J\}&=&2\sqrt{2}\delta^{IJ}P^-,\\
\{Q^+_I,Q^-_J\}&=&2Z_{IJ}.
\eea
As we mentioned before, in DLCQ diagonalization of $P^+$ is trivial and
the construction of Hamiltonian is the main problem. The last set of
equations suggests an alternative way of dealing with this problem: one
can first construct the matrix representation for the supercharge $Q^-$
and then just square it. This version of DLCQ first suggested in
\cite{sakai} appeared to be very fruitful. First of all it preserves
supersymmetry at finite resolution, while the conventional DLCQ
applied to supersymmetric theories doesn't.  The supersymmetric
version of DLCQ (SDLCQ) also provides the better numerical convergence.

To summarize,  we have two procedures for studying the
bound state spectrum: DLCQ and SDLCQ. To implement the first one we should
construct the light cone Hamiltonian and diagonalize it, while the second
approach requires the construction of the supercharge. Of course the SDLCQ
method is appropriate only for the theories with supersymmetries, although it
can be modified to study models with soft supersymmetry breaking.

\subsection{Reduction from Three Dimensions.}

Let us start by the defining a simple supersymmetric system in two
dimensions. It can be constructed by dimensional reduction of SYM from
three to two dimension.

Our starting point is the action for SYM in $2+1$ dimensions:
\be\label{3daction}
S=\int d^3 x \mbox{tr} \left(-\frac{1}{4} F_{AB}F^{AB}
+i \bar{\Psi}\gamma^A D_{A}\Psi \right).
\ee
The system consists of gauge field $A_A$ and two--component Majorana fermion
$\Psi$, both transforming according to adjoint representation of gauge group.
We assume that this group is either $U(N)$ or $SU(N)$ and thus matrices
$A^A_{ij}$ and $\Psi_{ij}$ are hermitian.
Studying dimensional reduction of $SYM_{D}$ we introduce the following
conventions for the indices: the capital latin letters correspond to $D$
dimensional spacetime, greek indices label two dimensional coordinates and
the lower case letters are used as matrix indices.
According to this conventions the indexes in (\ref{3daction}) go from
zero to two, the field strength $F_{AB}$ and covariant derivative $D_{A}$ are
defined in the usual way:
\bea
F_{AB}=\d_A A_B-\d_B A_A +ig[A_A,A_B],\nonumber\\
D_A\Psi=\d_A\Psi +ig[A_A,\Psi].
\eea

Dimensional reduction to $1+1$ means that we require all fields to be
independent on coordinate $x^2$, in other words we place the system on
the cylinder with radius $L_\perp$ along the $x^2$ axis and consider only
zero modes of the fields. We consider this reduction as a
formal
way of getting a two dimensional matrix model. In the reduced theory it is
convenient to introduce two dimensional indices and treat the  $A^2$
component of gauge field as a two dimensional scalar $\phi$. The action for
the reduced theory has the form:
\begin{equation}\label{2dact3dfer}
S  =  \int d^2 x \hspace{1mm}
 \mbox{tr} \left(-\frac{1}{4} F_{\mu \nu}F^{\mu \nu}
+\frac{1}{2}D_\mu \phi D^\mu \phi +
i \bar{\Psi}\gamma^\mu D_{\mu}\Psi -2ig\phi
\bar{\Psi}\gamma_5\Psi \right),
\end{equation}
We choose the
special representation of three dimensional gamma matrices:
\be
\gamma^0=\sigma^2,\qquad \gamma^1=i\sigma^1, \qquad \gamma^2=i\sigma^3,
\ee
then it is natural to write the spinor $\Psi$ in terms of its
components:
\be\label{3ddecomp}
\Psi=(\psi,\chi)^T.
\ee
Taking all these definitions into account one can rewrite the dimensional
reduction of (\ref{3daction}) as:
\bea\label{2daction}
S&=&L_\perp\int d^2x\left(\frac{1}{2}D_\mu\phi D^\mu\phi +i\sqrt{2}\psi D_+\psi
+i\sqrt{2}\chi D_-\chi+\right.\nonumber\\
&+&\left.2g\psi\{\psi,\chi\}-\frac{1}{4}F_{\mu\nu}F^{\mu\nu}\right).
\eea
The covariant derivatives here are taken with respect to the light cone
coordinates:
\be
x^\pm=\frac{x^0\pm x^1}{\sqrt{2}}.
\ee
Note that by rescaling the fields and coupling constant $g$ we can take the
constant $L_\perp$ to be equal to one.

The bound state problem for the system (\ref{2daction}) was first studied in
\cite{sakai}. The supersymmetric version of the discrete light cone
quantization was used in order to find the mass spectrum, and the zero modes
were neglected \cite{sakai}. We have found that while zero
modes are not very important for calculations of massive spectrum, they play
crucial role in the description of the vacuum of the theory.

Let us consider (\ref{2daction}) as the theory in the continuum. In
this case one can choose the light cone gauge:
\be
A^+=0,
\ee
then equations of motion for $A^-$ and $\chi$ give constraints:
\bea
&&-\partial_-^2 A^- =gJ^+, \\
&&\sqrt{2} i \partial_- \chi=g[\phi, \psi],\\
&&J^+(x)=\frac{1}{i}[\phi(x), \partial_-\phi(x)]-
\frac{1}{\sqrt 2}\{\psi(x), \psi(x)\}.
\eea
Solving this constraints and substituting the result back into the action
one determines the Lagrangian as function of physical fields $\phi$ and
$\psi$ only. Then using the usual Noether technique, we can construct the
conserved charges corresponding to the translational invariance:
\bea
P^+&=&\int dx^- \mbox{tr}\left((\partial_-\phi)^2+
    i\sqrt{2}\psi\partial_-\psi\right),\\
P^-&=&
 \int dx^-{\rm tr} \left(
-\frac{g^2}{2} J^+\frac{1}{\partial_-^2} J^+
+\frac{ig^2}{2\sqrt 2}[\phi, \psi]
\frac{1}{\partial_-}[\phi, \psi]\right).
\eea
We can also construct the Noether charges corresponding to the supersymmetry
transformation. However the naive SUSY transformations break the gauge fixing
condition $A^+=0$, so they should be accompanied by compensating a gauge
transformation:
\bea\label{susytrans}
\delta A_\mu=\frac{i}{2} {\bar\varepsilon}\gamma_\mu \Psi-D_\mu
\frac{i}{2} {\bar\varepsilon}\gamma_-\frac{1}{\partial_-} {\Psi},\\
\delta\Psi=\frac{1}{4}F_{\mu\nu}\gamma^{\mu\nu}\varepsilon-\frac{g}{2}[
{\bar\varepsilon}\gamma_-\frac{1}{\partial_-} {\Psi},\Psi].\nonumber
\end{eqnarray}
The resulting supercharges are:
\bea
Q^+=2\int dx^-\mbox{tr}\left(\psi\partial_-\phi\right),\\
Q^-=-2g\int dx^-\mbox{tr}\left(J^+\frac{1}{\partial_-}\psi\right).
\eea

Consider now the general reduction of SYM$_D$ to two
dimensions. By counting the fermionic and bosonic degrees of freedom
one can see that the SYM can be defined only in a limited number of
spacetime dimensions, namely $D$ can be equal to 2, 3, 4, 6 or 10. The
last case is the most general one: all other system can be obtained by
dimensional reduction and appropriate truncation of degrees of freedom.
So we will concentrate on the reduction $10\rightarrow 2$, and
the comments on four and six dimensional cases will be made in the end.

As in the last subsection we start from ten dimensional action:
\be\label{10daction}
S=\int d^3 x \mbox{tr} \left(-\frac{1}{4} F_{AB}F^{AB}+
i \bar{\Psi}\gamma^A D_{A}\Psi \right).
\ee

According to our general conventions the indexes in (\ref{10daction}) go from
zero to nine, $\Psi$ is the ten dimensional Majorana--Weyl spinor. A general
spinor in ten dimensions has $2^{10/2}=32$ complex components, if the
appropriate basis of gamma matrices is chosen then the Majorana condition makes
all the components real. Since all the matrices in such representation are
real, the Weyl condition
\begin{equation}
\Gamma_{11}\Psi=\Psi
\end{equation}
is compatible with the reality of $\Psi$ and thus it eliminates half of its
components. In the special representation of Dirac matrices:
\begin{eqnarray}
&& \Gamma^0=\sigma_2 \otimes {\bf 1}_{16}, \\
&& \Gamma^I={\rm i}\sigma_1 \otimes \gamma^I, \hspace{6mm} I=1,\dots,8;\\
&& \Gamma^9= {\rm i}\sigma_1 \otimes \gamma^9,
\end{eqnarray}
the $\Gamma_{11}= \Gamma^0 \cdots \Gamma^9$ has very simple form:
$\Gamma_{11}=\sigma_3\otimes {\bf 1}_{16}$. Then the Majorana spinor of
positive chirality can be written in terms of 16--component real object
$\psi$:
\begin{equation}
\Psi= 2^{1/2} { \psi \choose 0}. \label{spin16}
\end{equation}

Let us return to the expressions for $\Gamma$ matrices. The ten dimensional
Dirac algebra
$$
\{\Gamma_\mu,\Gamma_\nu\}=2g_{\mu\nu}
$$
is equivalent to the spin(8) algebra for $\gamma$ matrices:
$\{\gamma_I,\gamma_J\}=2\delta_{IJ}$ and the ninth matrix can be chosen to
be $\gamma^9=\gamma^1\dots\gamma^8$. Note that the 16 dimensional
representation of spin(8) is the reducible one: it can be decomposed as
${\bf 8}_s+{\bf 8}_c$
\begin{equation}
\gamma^I=\left(\begin{array}{cc}
0 & \beta_I\\
\beta_I^T & 0
\end{array}\right), \hspace{7mm} I=1,\dots,8.
\end{equation}
The explicit expressions for the $\beta_I$ satisfying
$\{\beta_I,\beta_J\}=2\delta_{IJ}$ can be found in \cite{GSW}. Such choice
leads to the convenient form of $\gamma^9$:
\begin{equation}
\gamma^9=\left(\begin{array}{cc}
{\bf 1}_{8} & 0\\
0 & -{\bf 1}_{8} \end{array}\right). \label{gamma9}
\end{equation}

So far we have found nonzero components of the spinor given by (\ref{spin16}).
However as we saw before not all such components are physical
in the light cone gauge, so it is useful to perform the analog of
decomposition (\ref{3ddecomp}). In ten dimension it is related with breaking
the sixteen component spinor $\psi$ on the left and right--moving components
using the projection operators
\begin{equation}
P_L=\frac{1}{2}(1-\gamma^9), \qquad P_R=\frac{1}{2}(1+\gamma^9).
\end{equation}
After introducing the light--cone coordinates
$x^\pm=\frac{1}{\sqrt{2}}(x^0\pm x^9)$ the action (\ref{10daction}) can be
rewritten as
\begin{eqnarray}
S_{9+1}^{LC} & = & \int dx^+ dx^- d{\bf x}^{\perp} \hspace{1mm}
 \mbox{tr} \left( \frac{1}{2}F_{+-}^2 + F_{+I}F_{-I} - \frac{1}{4}F_{IJ}^2
 \right. \nonumber \\
& & \hspace{20mm}
+ \hspace{1mm}
i\sqrt{2} \psi_R^T D_+ \psi_R + i\sqrt{2}\psi_L^T D_- \psi_L +
     2i\psi_L^T \gamma^I D_I \psi_R \left.
\frac{}{} \right),
\label{LCversion}
\end{eqnarray}
where the repeated indices $I,J$ are summed over $(1,\dots,8)$. After applying
the light--cone gauge $A^+=0$ one can eliminate nonphysical degrees of freedom
using the Euler--Lagrange equations for $\psi_L$ and $A^-$:
\begin{eqnarray}
\label{fermioncon}
 \partial_- \psi_L = -\frac{1}{\sqrt{2}}\gamma^I D_I \psi_R, \\
\label{apluscon}
\partial_{-}^2 A_{+}=\partial_{-}\partial_{I}A_{I}+gJ^+\\
J^+=i[A_{I},\partial_{-}A_{I}]+2\sqrt{2}\psi_{R}^T\psi_{R}.
\end{eqnarray}

Performing the reduction to two dimensions means that all fields are assumed
to be independent on the transverse coordinates: $\partial_{I}\Phi=0$. Then
as before one can construct the conserved momenta $P^\pm$ in
terms of physical degrees of freedom:
\begin{eqnarray}
P^+&=&\int dx^- \mbox{tr}\left((\partial_-A_I)^2+
    i\sqrt{2}\psi_R\partial_-\psi_R\right),\\
P^-&=&
 \int dx^-{\rm tr} \left(
-\frac{g^2}{2} J^+\frac{1}{\partial_-^2} J^+
+\frac{ig^2}{2\sqrt 2}[A_I, \psi_R^T] \beta^T_I
\frac{1}{\partial_-}\beta_J[A_J, \psi_R]\right)-\nonumber\\
&-&\frac{1}{4}\int dx^-{\rm tr}\left([A_IA_J]^2\right).
\end{eqnarray}

We can also construct the Noether charges corresponding to the supersymmetry
transformation (\ref{susytrans}). As in the three dimensional case it is
convenient to decompose the supercharge in two components:
$$
Q^+=P_L Q,\qquad Q^-=P_R Q.
$$
The resulting eight component supercharges are given by
\begin{eqnarray}
Q^+=2\int dx^-\mbox{tr}\left(\beta^T_I\psi_R\partial_-A_I\right),\\
Q^-=-2g\int dx^-\mbox{tr}\left(J^+\frac{1}{\partial_-}\psi_R+
\frac{i}{4}[A_IA_J](\beta_I\beta^T_J-\beta_J\beta^T_I)\psi_R\right).
\end{eqnarray}

Finally we make a short comment on dimensional reduction of $SYM_{3+1}$ and
$SYM_{5+1}$. These systems can be constructed repeating the procedure just
described. However there is an easier way to construct the Hamiltonian and
supercharges for the dimensionally reduced theories, namely one has to
truncate the unwanted degrees of freedom in the ten dimensional expressions.
This is especially easy for the bosonic coordinates: one simply considers
indices $I$ and $J$ running from one to two (for $D=4$) or to four (for $D=6$).
The fermionic truncation can also be performed by requiring the spinor
$\psi_R$ to be 2-- or 4--component. Then the only problem is the choice of
$2\times 2$ or $4\times 4$ beta matrices satisfying
\begin{equation}
\{\beta_I,\beta_J\}=2\delta_{IJ},
\end{equation}
which is an easy task.

\section*{Conclusion.}

While we are still far from completely solving the bound
state problem in three and four dimensional theories, we can
already  make some statements about these theories:

We have studied the structure of bound states for two dimensional
supersymmetric models  obtained by dimensional reduction from SYM$_{2+1}$.
For this theory we have proven that any normalizable bound state in the
continuum must include a contribution with arbitrarily large number of
partons and we have shown that this is the general property of supersymmetric
matrix models \cite{alp98}. This scenario is to be contrasted with the simple
bound states discovered in a number of, $1+1$ dimensional theories with
complex fermions, such as the Schwinger model, the t`Hooft model, and a
dimensionally reduced theory with complex adjoint fermions
\cite{anp97,pin97}. We also study the massless states of SYM$_{2+1}$ in
DLCQ. Some of them are constructed explicitly and the general formula for
the number of massless states as function of harmonic resolution is derived
for the large $N$ case \cite{alp98a}.

Some of our results on zero modes in 2 dimensions are discussed in
\cite{alptzm}.
They can be used to describe the vacuum structure of
$SYM_{2+1}$ on a cylinder because only zero modes contributes to such
structures. Thus studying the dimensionally reduced theory in $1+1$ provides
all the necessary information. We found that two dimensional models also
determine the behavior of bound states at weak coupling in three dimensions
and determine the exact number of massless states. We performed such counting
only for (1,1) theory. The theory with (2,2) supersymmetry theory
\cite{appt} and
the even more interesting case of the (8,8) theory
\cite{alppt}, which is known to have a mass gap have not yet been addressed.

The bound state problem we have studied so far is the traditional one
for DLCQ. However this is not the only calculation that can be done us
this method. The problem of computing of correlation functions, more
traditional for conventional quantum field theory, can also be addressed
in the light cone quantization. Unlike the usual methods of QFT, DLCQ
calculations are valid beyond perturbation theory and thus can be used
for testing the duality between gauge theory and supergravity.

There has been a great deal of excitement during this past year
following the realization that certain field theories admit concrete
realizations as a string theory on a particular background
\cite{adscft}. By now many examples of this type of correspondence for
field theories in various dimensions with various field contents have
been reported in the literature (for a comprehensive review and list
of references, see \cite{agmoo}).  However, attempts to apply these
correspondences to study the details of these theories have only met
with limited success so far. The problem stems from the fact that our
understanding of both sides of the correspondence is limited. On the
field theory side, most of what we know comes from perturbation theory
where we assume that the coupling is weak. On the string theory side,
most of what we know comes from the supergravity approximation where
the curvature is small.  There are no known situations where both
approximations are simultaneously valid. At the present time,
comparisons between the dual gauge/string theories have been
restricted to either qualitative issues or quantities constrained by
symmetry. Any improvement in our understanding of field theories
beyond perturbation theory or string theories beyond the supergravity
approximation is therefore a welcome development.

We have studied the field theory/string theory correspondence \cite{ahlp99}
motivated by considering the near-horizon decoupling limit of a D1-brane in
type
IIB string theory \cite{IMSY}. The gauge theory corresponding to this
theory is the Yang-Mills theory in two dimensions with 16
supercharges.  Its SDLCQ formulation was recently reported in
\cite{alppt}. This is probably the simplest known example of a field
theory/string theory correspondence involving a field theory in two
dimensions with a concrete Lagrangian formulation.

A convenient quantity that can be computed on both sides of the
correspondence is the correlation function of gauge invariant
operators \cite{GKP,Wit}. We  focused on two point functions of
the stress-energy tensor.  This turns out to be a very convenient quantity
to compute for many reasons.  Some
aspects of this as it pertains to a consideration of black hole entropy
was recently discussed in \cite{akisunny}. There are other physical
quantities often reported in the literature. In the DLCQ literature,
the spectrum of hadrons is often reported.  This would be fine for
theories in a confining phase. However, we expect the SYM in two
dimension to flow to a non-trivial conformal fixed point in the
infra-red \cite{IMSY,DVV}.  The spectrum of states will therefore form
a continuum and will be cumbersome to handle.  On the string theory
side, entropy density \cite{BISY} and the quark anti-quark potential
\cite{BISY,RY,juanwilson} are frequently reported. The definition of
entropy density requires that we place the field theory in a
space-like box which seems incommensurate with the discretized light
cone.  Similarly, a static quark anti-quark configuration does not fit
very well inside a discretized light-cone geometry.  The correlation
function of point-like operators do not suffer from these problems.

Let us now mention the immediate challenges to DLCQ following from our
consideration. First of all it is straightforward to extend the numerical
results for the correlator to higher resolution and thus to test the
Maldacena conjecture. The only problem here is the limits in one's
computing resources. The better computer power may also help to extend our
analysis of three dimensional system to larger values of transverse
truncation and it might be possible to extrapolate the results to
continuum. The simple transverse truncation we have used so far do not
provide much information about behavior of the spectrum as function of
transverse resolution. Another serious disadvantage of our approach is that
we study the theory on the cylinder and thus the structure of the
topological excitations of our system differs from the one of SYM$_{2+1}$ in
infinite spacetime. Such excitations can become important in the strong
coupling limit. However, even the spectrum for the theory on the cylinder is
of a great interest and we leave the detailed numerical study of this system
for future work. Finally solving for bound states of four dimensional
theories is still the greatest challenge for DLCQ.

In this talk we have reviewed some of the progress in the application of
discrete light cone quantization to the supersymmetric systems. Studying such
systems is especially interesting because the cancellation between bosonic
and fermionic loops make these theories much easier to renormalize than
models without supersymmetry. Although we didn't need this advantage when
considering two dimensional systems, it becomes crucial in higher
dimensions. From this point of view it is desirable to have exact SUSY  in
discretized theories to simplify the renormalization in DLCQ.

\section{Acknowledgments}
This work was supported in part by the US Department of Energy.


\begin{thebibliography}{9}

\bibitem{agmoo}
O.~Aharony, S.~S. Gubser, J.~Maldacena, H.~Ooguri, and Y.~Oz, ``Large N field
  theories, string theory and gravity,''
  {{\tt hep-th/9905111}}.

\bibitem{alppt}
F.~Antonuccio, O.~Lunin, S.~Pinsky, H.~C. Pauli, and S.~Tsujimaru, ``The DLCQ
  spectrum of N=(8,8) superYang-Mills,'' {\em Phys. Rev.} {\bf D58} (1998)
  105024, {{\tt hep-th/9806133}}.

\bibitem{alptzm}
F.~Antonuccio, O.~Lunin, S.~Pinsky, and S.~Tsujimaru, ``The Light cone vacuum
  in (1+1)-dimensional superYang-Mills theory,''
  {{\tt hep-th/9811254}}.

\bibitem{appt}
F.~Antonuccio, H.~C. Pauli, S.~Pinsky, and S.~Tsujimaru, ``DLCQ bound states of
  N=(2,2) super Yang-Mills at finite and large N,'' {\em Phys. Rev.} {\bf D58}
  (1998) 125006, {{\tt hep-th/9808120}}.

\bibitem{anp97} F.~Antonuccio, S.S.~Pinsky,
 "Matrix theories from reduced $SU(N)$ Yang--Mills with adjoint fermions,"
 {\em Phys.Lett} {\bf B397} (1997) 42, {{\tt hep-th/9612021}}.

\bibitem{bdk93} G.~Bhanot, K.~Demeterfi, I.R.~Klebanov,
"$(1+1)$--Dimensional Large  N QCD Coupled to Adjoint fermions,"
{\em Phys. Rev.} {\bf D48} (1993) 4980, {{\tt hep--th/9307111}}.

\bibitem{kub94} J.~Boorstein, D.~Kutasov,
 "Symmetries and mass splittings in QCD$_2$ coupled to adjoint fermions,"
  {\em Nucl.Phys.} {\bf B421} (1994) 263, {{\tt hep-th/9401044}}.

\bibitem{BISY}
A.~Brandhuber, N.~Itzhaki, J.~Sonnenschein, and S.~Yankielowicz, ``Wilson
  loops, confinement, and phase transitions in large N gauge theories from
  supergravity,'' {\em JHEP} {\bf 06} (1998) 001,
  {{\tt hep-th/9803263}}.

\bibitem{BPP}
S.~J. Brodsky, H.-C. Pauli, and S.~S. Pinsky, ``Quantum chromodynamics and
  other field theories on the light cone,'' {\em Phys. Rept.} {\bf 301} (1998)
  299, {{\tt hep-ph/9705477}}.

\bibitem{DVV}
R.~Dijkgraaf, E.~Verlinde, and H.~Verlinde, ``Matrix string theory,'' {\em
  Nucl. Phys.} {\bf B500} (1997) 43,
  {{\tt hep-th/9703030}}.

\bibitem{Dirac}
P.~A.~M. Dirac, ``Forms of relativistic dynamics,'' {\em Rev. Mod. Phys.} {\bf
  21} (1949) 392.

\bibitem{givkut} A.~Giveon, D.~Kutasov,
 "Brane dynamics and gauge theory,"
 {{\tt hep-th/9802067}}.

\bibitem{GKP}
S.~S. Gubser, I.~R. Klebanov, and A.~M. Polyakov, ``Gauge theory correlators
  from noncritical string theory,'' {\em Phys. Lett.} {\bf B428} (1998) 105,
  {{\tt hep-th/9802109}}.



\bibitem{akisunny}
A.~Hashimoto and N.~Itzhaki, ``A Comment on the Zamolodchikov c function and
  the black string entropy,''
  {{\tt hep-th/9903067}}.

\bibitem{IMSY}
N.~Itzhaki, J.~M. Maldacena, J.~Sonnenschein, and S.~Yankielowicz,
  ``Supergravity and the large N limit of theories with sixteen supercharges,''
  {\em Phys. Rev.} {\bf D58} (1998) 046004,
  {{\tt hep-th/9802042}}.

\bibitem{kut93} D.~Kutasov,
"Two Dimensional QCD Coupled to Adjoint Matter and String Theory,"
{\em Phys. Rev.} {\bf D48} (1993) 4980,  {{\tt hep--th/9306013}}.

\bibitem{adscft}
J.~Maldacena, ``The Large N limit of superconformal field theories and
  supergravity,'' {\em Adv. Theor. Math. Phys.} {\bf 2} (1998) 231,
  {{\tt hep-th/9711200}}.

\bibitem{juanwilson}
J.~Maldacena, ``Wilson loops in large N field theories,'' {\em Phys. Rev.
  Lett.} {\bf 80} (1998) 4859,
  {{\tt hep-th/9803002}}.

\bibitem{MaYam}
T.~Maskawa and K.~Yamawaki, ``The problem of $p^+ = 0$ mode in the null plane
  field theory and Dirac's method of quantization,'' {\em Prog. Theor. Phys.}
  {\bf 56} (1976) 270.

\bibitem{sakai}
Y.~Matsumura, N.~Sakai, and T.~Sakai, ``Mass spectra of supersymmetric
  Yang-Mills theories in (1+1)-dimensions,'' {\em Phys. Rev.} {\bf D52} (1995)
  2446--2461, {{\tt hep-th/9504150}}.

\bibitem{BP85}
H.~C. Pauli and S.~J. Brodsky, ``Discretized light cone quantization: solution
  to a field theory in one space one time dimensions,'' {\em Phys. Rev.} {\bf
  D32} (1985) 1993, 2001.

\bibitem{pin97}
S.~Pinsky,
 "The Analog of the t'Hooft Pion with Adjoint Fermions,"
  Invited talk at New Nonperturbative Methods and Quantization of the Light
  Cone, Les Houches, France, 24 Feb - 7 Mar 1997. {{\tt hep-th/9705242}}.

\bibitem{RY}
S.-J.~Rey and J.~Yee, ``Macroscopic strings as heavy quarks in large N gauge
  theory and anti-de Sitter supergravity,''
  {{\tt hep-th/9803001}}.


\bibitem{seiberg} N.~Seiberg,
 "Electric-magnetic duality in supersymmetric non-Abelian gauge theories,"
 {\em Nucl.Phys.} {\bf B435} (1995) 129

\bibitem{seibergwitten} N.~Seiberg, E.~Witten,
 "Monopoles, duality and chiral symmetry breaking in N=2 supersymmetric
    QCD,"
 {\em Nucl.Phys.} {\bf B431} (1994) 484.

\bibitem{WessBag} J.~Wess, J.~Bagger,
 {\em Supersymmetry and supergravity}, Princeton University Press(1992).

\bibitem{ahlp99} F.~Antonuccio, A.~Hashimoto, O.~Lunin, and S.~Pinsky, {\em
JHEP}
{\bf 9907:029, 1999}, {{\tt hep-th/9906087}}

\bibitem{witt96} E.~Witten,
 "Bound states of strings and p--branes,"
 {\em Nucl.Phys.} {\bf B460}, (1996) 335, {{\tt hep-th/9510135}}.

\bibitem{Wit}
E.~Witten, ``Anti-de Sitter space and holography,'' {\em Adv. Theor. Math.
  Phys.} {\bf 2} (1998) 253, {{\tt hep-th/9802150}}.

\bibitem{alp98}
F.~Antonuccio, O.~Lunin, and S.~Pinsky, ``Bound States of Dimensionally
Reduced SYM(2+1) at Finite N'' Phys. Lett. B429: 327-335, 1998.
  {{\tt hep-th/9803027}}.

\bibitem{alp98a}
F.~Antonuccio, O.~Lunin, and S.~Pinsky, ``Nonperturbative Spectrum of
Two-Dimensional (1,1) Superyand-Mills at Finite and Large N'' Phys. Rev. D58:
085009, 1998.
  {{\tt hep-th/9803170}}.

\end{thebibliography}
\end{document}